\documentclass[runningheads]{svmult}

\usepackage{makeidx}   % allows index generation
\usepackage{graphicx}  % standard LaTeX graphics tool
\usepackage{subeqnar}  % subnumbers individual equations
\usepackage{multicol}  % used for the two-column index
\usepackage{physprbb}  % modified textarea for proceedings,
\makeindex             % used for the subject index
\begin{document}
\title*{Infrared Spectroscopy of Brown Dwarfs:\protect\newline the onset
of CH$_{4}$ absorption in L dwarfs and the L/T transition}
\toctitle{Infrared Spectroscopy of Brown Dwarfs:\protect\newline the
onsetof CH$_{4}$ absorption in L dwarfs and the L/T transition}
% allows explicit linebreak for the table of content
%
%
\titlerunning{IR Spectroscopy of Brown Dwarfs}
% allows abbreviation of title, if the full title is too long
% to fit in the running head
%
\author{T. R. Geballe\inst{1}
\and K. S. Noll\inst{2}
\and S. K. Leggett\inst{3}
\and G. R. Knapp\inst{4}
\and X. Fan\inst{4}
\and D. Golimowski\inst{5}}

\authorrunning{T. R. Geballe et al.}
% if there are more than two authors,
% please abbreviate author list for running head
%
%
\institute{Gemini Observatory, 670 N. A'ohoku Pl., Hilo, HI 96720 USA
\and Space Telescope Science Institute, 3700 San Martin Dr., Baltimore, MD
21218 USA
\and Joint Astronomy Centre, 660 N. A'ohoku Pl., Hilo, HI 96720 USA
\and Princeton University Observatory, Princeton, NJ 08544 USA
\and Dept. of Physics and Astronomy, Johns Hopkins Univ., 3701 San Martin
Dr., Baltimore, MD 21218 USA
}

\maketitle              % typesets the title of the contribution

\begin{abstract}

We present infrared spectra of brown dwarfs with spectral types from
mid-L to T. The 0.9-2.5~$\mu$m spectra of three dwarfs found by the Sloan
Digital Sky Survey contain absorption bands of both methane and carbon
monoxide and bridge the gap between late L and previously observed T
dwarfs. These dwarfs form a clear spectral sequence, with CH$_{4}$
absorption increasing as the CO absorption decreases. Water vapor band
strengths increase in parallel with the methane bands and thus also link
the L and T types. We suggest that objects with detectable CO and
CH$_{4}$ in the H and K bands should define the earliest T subclasses.
From observations of bright (K~$\leq$~13~mag) L dwarfs found by 2MASS, we
find that the onset of detectable amounts of CH$_{4}$ occurs near
spectral type L5. For this spectral type methane is observable in the
3.3~$\mu$m $\nu$$_{3}$ band only, and not in the overtone and combination
bands at H and K. 

\end{abstract}

\section{Introduction}  

The infrared spectra of T type brown dwarfs are profoundly affected by
the molecules methane (CH$_{4}$) and water (H$_{2}$O). Huge bites are
taken out of the spectra by absorption bands of these molecules. At other
infrared wavelengths, adjacent to some of these absorption bands, the
atmospheres of the dwarfs are remarkably transparent. A photosphere with
an effective temperature of 950~K, the value for Gliese~229B, might
naively be expected to emit its maximum flux density near the peak of a
blackbody of that temperature, i.e., just longward of 3~$\mu$m. However,
as is shown in Fig. 1 (upper panel) the maximum for Gl~229B occurs in the
J band, near 1.2~$\mu$m, almost a factor of three shorter wavelength. The
strongest band of CH$_{4}$ is in fact centered at 3.3~$\mu$m (very close
to the peak of a 950~K black body) and virtually no radiation emerges
from the brown dwarf near that wavelength. In the J, H, and K bands the
locations of the CH$_{4}$ and H$_{2}$O absorptions give a T dwarf roughly
the JHK colors of a very hot star.

\begin{figure}
\centerline{\includegraphics[width=32pc]{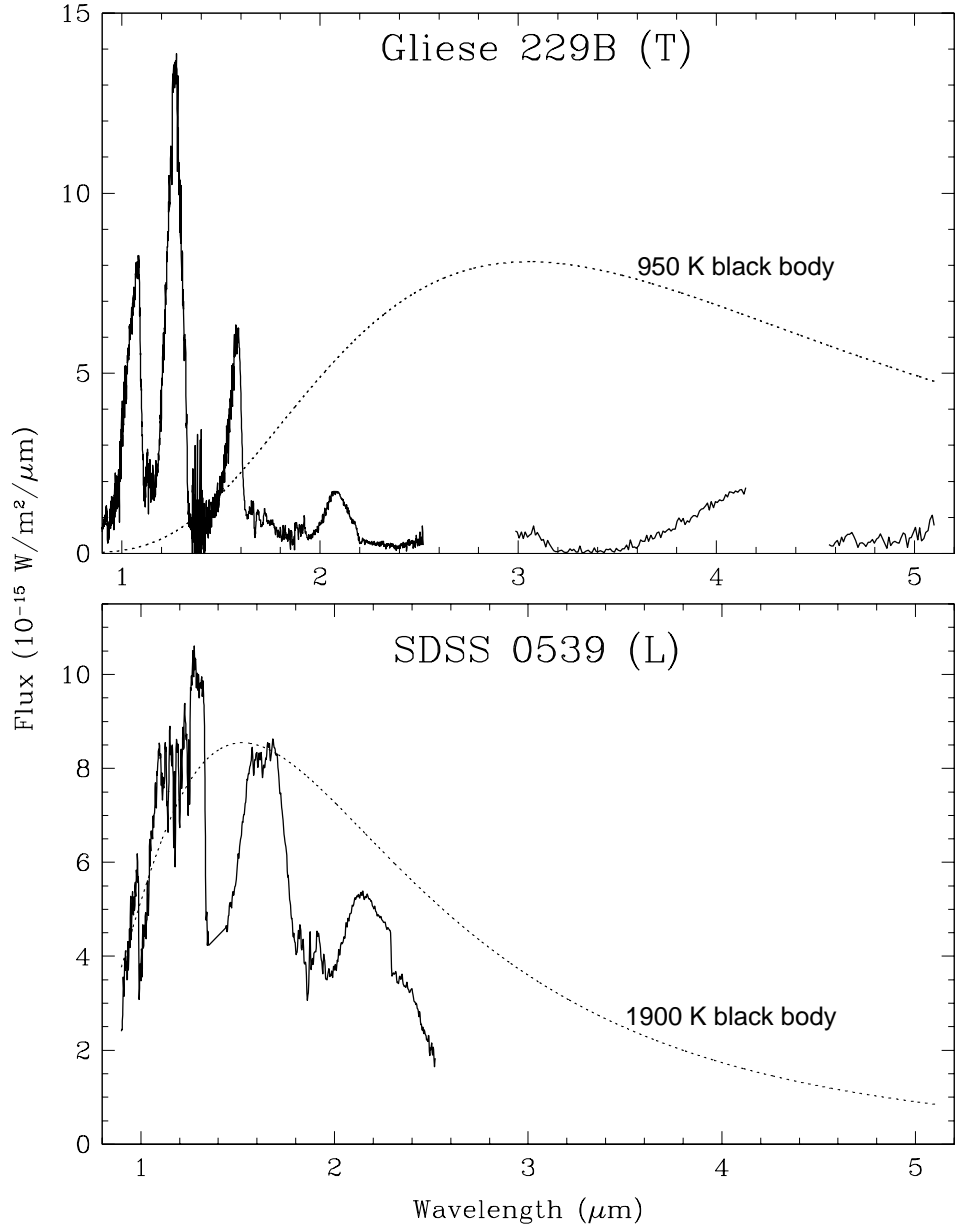}}
\caption{Spectra of representative T and L type brown dwarfs, Gl~229B
[1,2,3] and SDSS~0539 [4] compared with black body functions (of arbitrary
strengths) corresponding to the effective temperatures of the brown
dwarfs. The original Gl~229B data have been recalibrated [5,6].}
\end{figure}

The infrared spectrum of an L dwarf (Fig.~1, lower panel) also is badly
eaten away by molecular absorptions, many of which are different than
those affecting a T dwarf. However, the maximum flux density occurs close
to 1.5~$\mu$m, the peak of a blackbody whose temperature is $\sim$2000~K,
roughly the effective temperature of the dwarf. In addition, the JHK
colors of L dwarfs are much closer to what one naively would expect for an
object of that temperature.

Thus in the key J, H, and K bands the spectral transition of a brown dwarf
from L to T is a huge change, much greater than between any other two
neighboring spectral types. The transformation is largely due to the
increasing stability of methane at lower temperatures, its increasing
abundance as the overwhelming presence of hydrogen drives the chemical
equilibrium of carbon-bearing species toward dominance by CH$_{4}$, and
the resulting onset of strong absorption by CH$_{4}$ in a variety of
combination and overtone bands in the 1-2.5~$\mu$m region.

The bands of CH$_{4}$ in the 1-2.5$\mu$m interval are strong in T dwarfs,
but do not correspond to fundamental vibrations of the molecule. Those
much stronger bands occur at longer wavelengths where, because of poor
atmospheric transmission and high sky and telescope backgrounds, few
spectra of L and T dwarfs have been obtained.

\section{T-Dwarf spectra before 2000}

Following the discovery of Gl~229B, it took five years before the Sloan
Digital Sky Survey (SDSS), the Two-Micron All Sky Survey (2MASS), and the
European Southern Observatory (ESO) found additional T dwarfs [7,8,9,10]. 
The 1-2.5~$\mu$m spectra of the objects they discovered are nearly
identical to that of Gl~229B.  There are in fact some differences between
them, probably related to differences in temperature and surface gravity,
but they are fairly subtle. At the end of 1999 no objects were known with
spectra located in the huge gap, demonstrated in Fig.~1, between the
latest L type, which showed no CH$_{4}$ absorption at 1-2.5~$\mu$m and
the T types, in which very strong CH$_{4}$ bands are present.  It was not
clear at the time whether this lack was an observational selection effect
or indicated that the transition from latest L-types to then known
T-types was rapid compared to the overall cooling time of an L dwarf,
encompassing only a relatively narrow temperature range.

\begin{figure}
\centerline{\includegraphics[width=32pc]{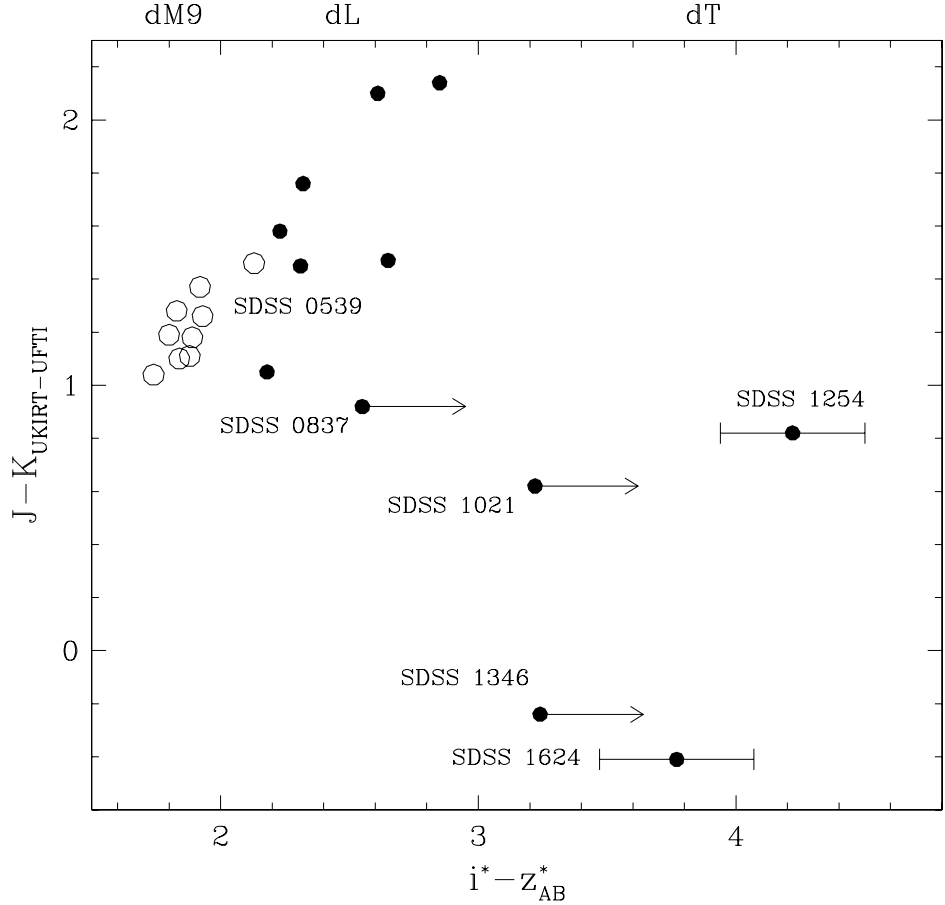}}
\caption{UKIRT(UFTI) J-K vs. SDSS $i-z$ [4]. The objects in the top left
quadrant are late M and L dwarfs; those at the bottom are classical T
dwarfs.}
\end{figure}

It now is evident that the lack of objects in the transition region was
largely observational selection. This is demonstrated in Fig.~2, in which
$J-K$ is plotted vs. $i-z$ for L and T dwarfs. The area at the top left
of the figure is the domain of the L dwarfs, the area at bottom right is
the domain of the classical T dwarfs. In 1999 the majority (5) of all (8)
published T dwarf identifications were from 2MASS, which surveys the sky
at J, H, and K {\it only}. In Fig.~2 it can be seen that the T dwarfs
have blue or bluer $J-K$ colors than all but the hottest stars. Objects
with such colors form a very small subset of the 2MASS catalogue.
Determining if such objects are candidate brown dwarfs is relatively
straightforward, requiring comparison with the Palomar Sky Survey.
However, if, as expected, during the transition between L and T, brown
dwarfs follow a direct path in Fig.~2 between these two domains, their
$J-K$ colors will pass through the same range of values as the myriads of
cool main sequence stars and they will be photometrically
indistinguishable from such stars by 2MASS.  In contrast, the transition
objects would be expected to develop even redder and more unusual $i-z$
colors and, as in the case of T dwarfs, to be readily singled out by
SDSS. However, by the end of 1999 the two objects at the bottom of Fig.~2
were the only SDSS objects known to be later than L.

\begin{figure}
\centerline{\includegraphics[width=32pc]{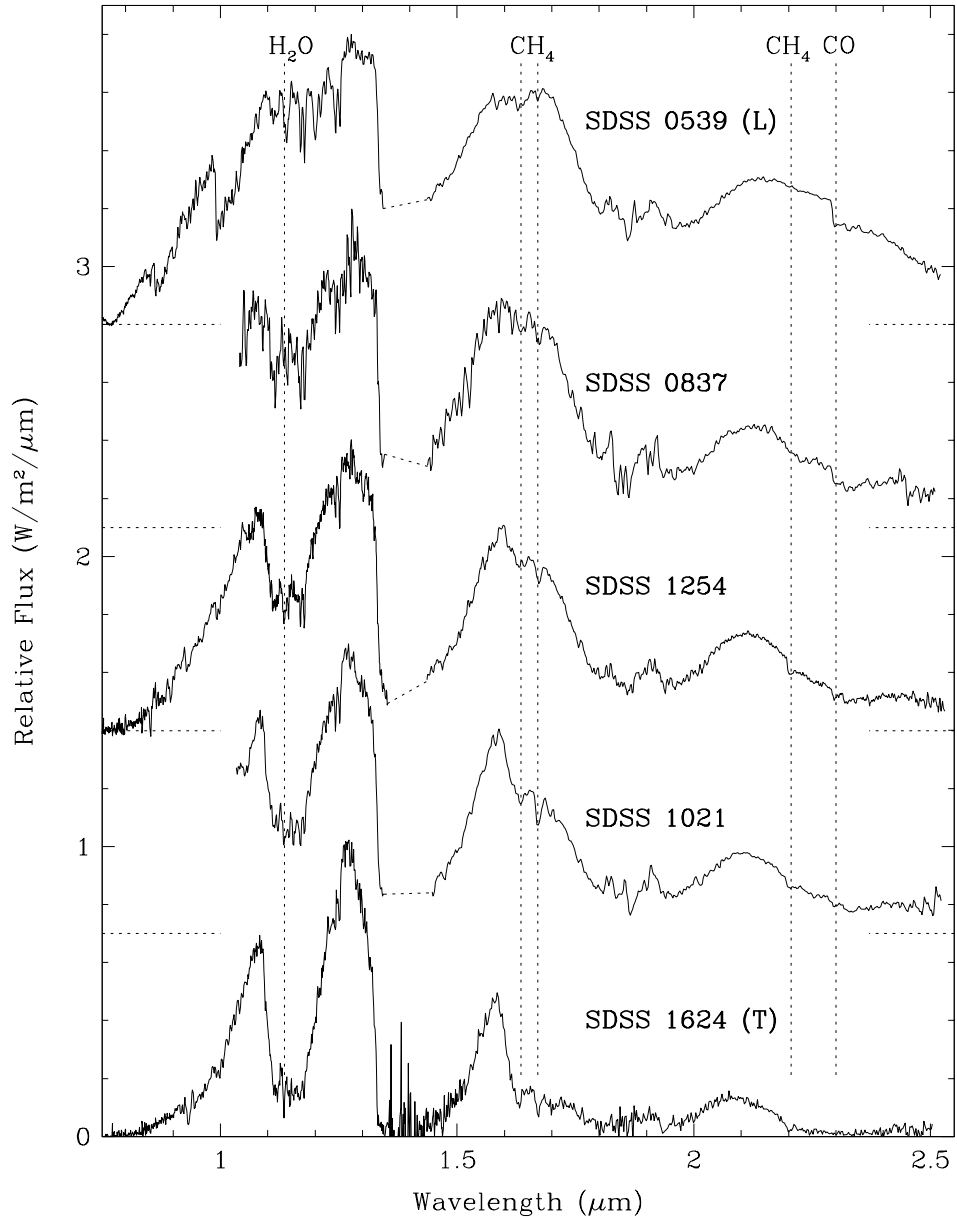}}
\caption{Spectra of three transition objects together with a mid-L dwarf
(top) and a classical T dwarf (bottom) [4]. The dashed vertical lines
mark the wavelengths of bands (or, in the case of CO, the 2-0 band
head) whose strengths change rapidly in the transition region. The
horizontal dashed lines are zero flux density levels for the individual
spectra.} 
\end{figure}

\section{Transition Objects}

In early 2000 three new objects were identified as candidate T dwarfs by
SDSS, on the basis of their $i-z$ colors.  JHK photometry at UKIRT [4]
showed that these objects (SDSS~0837, SDSS~1021, and SDSS~1254) inhabited
the region in Fig.~2 between the previously known L and T dwarfs. Spectra
of them were obtained at the United Kingdom Infrared Telescope (UKIRT) in
late February and March, 2000 and are displayed in Fig.~3, together with
those of a mid-L dwarf and a previously known SDSS T dwarf.  These three
spectra fortuitously delineate the transition between L and T fairly
evenly, with CO overtone bands at 2.3-2.4~$\mu$m gradually disappearing
from top to bottom in the figure and the methane bands at 1.6-1.8~$\mu$m
and 2.2-2.5~$\mu$m gradually strengthening. One of the largest spectral
changes is in the strength of the water band ($\nu_{1}+\nu_{2}+\nu_{3}$)
at 1.15~$\mu$m. That band is weak in the L dwarf at the top of Fig.~3,
but increases steadily in the transition objects, and is nearly totally
absorbing in the classical T dwarf.

At the time of this conference, SDSS has discovered comparable numbers of
classical T dwarfs and transition objects. Although the numbers of each
are few, they suggest that both transition dwarfs and classical T dwarfs
occupy substantial temperature ranges. Hence it is likely that many more
of each kind will be found by SDSS.

\begin{figure}
\centerline{\includegraphics[width=32pc]{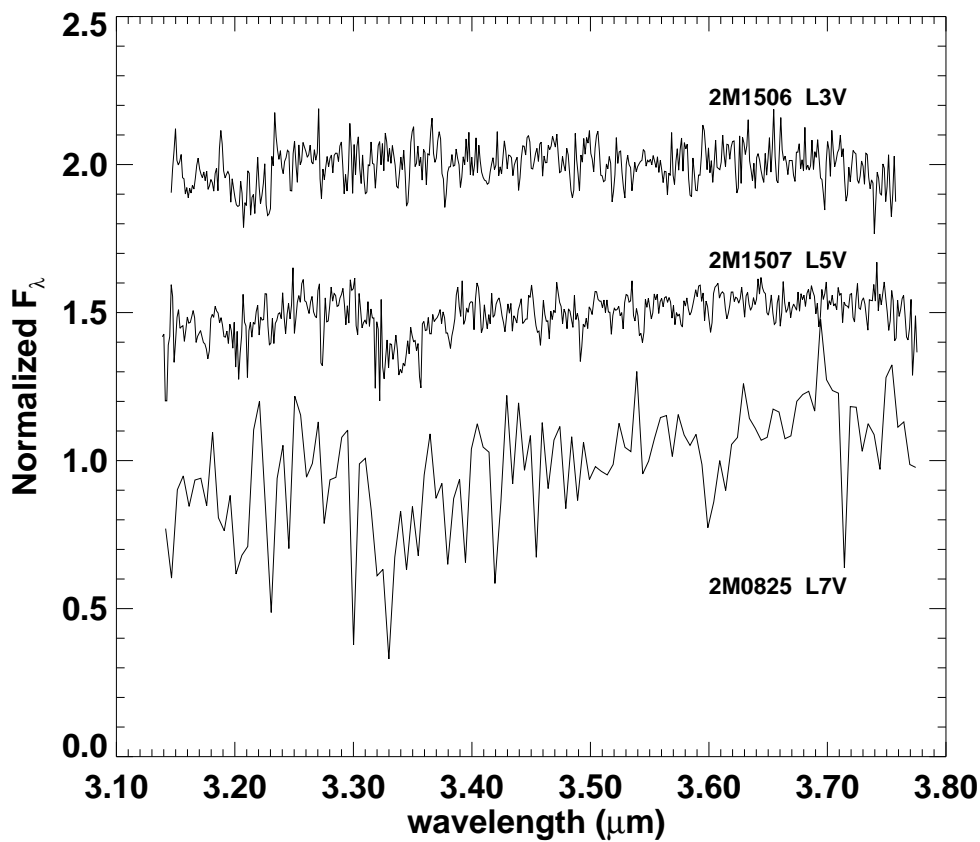}}
\caption{Spectra of three L-type brown dwarfs in the region of the methane
$\nu_{3}$ fundamental band [11]. The broad absorption present in the two
later type dwarfs is the methane Q-branch.}
\end{figure}

\begin{figure}
\centerline{\includegraphics[width=32pc]{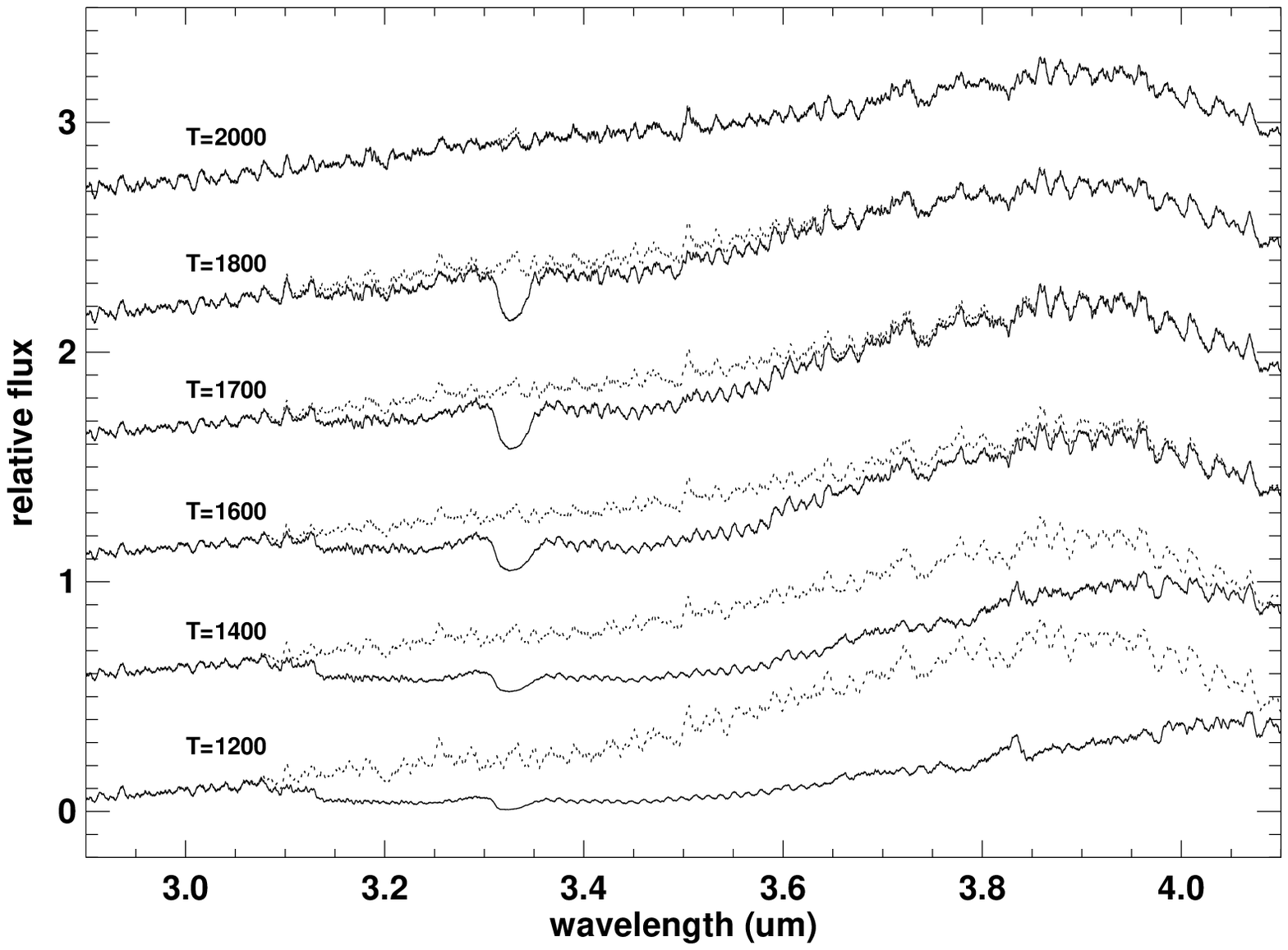}}
\caption{Model 2.9-4.1~$\mu$m spectra for dwarfs with effective
temperatures ranging from 1200~K to 2000~K.  Dashed curves contain no
methane, and the spectral features are primarily due to water vapor. Solid
curves have absorption from both H$_2$O and CH$_4$. The $\nu_3$ Q branch
first becomes detectable at T$_{eff} \approx$ 1900~K.  The P and R
branches begin to significantly suppress the surrounding H$_2$O
pseudo-continuum by T$_{eff} \approx$ 1700~K.}
\end{figure}

\section{Methane absorption in L dwarfs}

As mentioned earlier, the most intense bands of methane occur outside the
JHK region, at longer wavelengths. The strongest of these is the
$\nu_{3}$ band centered at 3.3~$\mu$m. This band, one-hundred or more
times stronger than the any of the CH$_{4}$ bands in the 1-2.5~$\mu$m
interval, ought to appear at higher temperatures than the 1-2.5~$\mu$m
bands and must already be very strong in the transition objects discussed
above. Thus, one would predict that in a late L dwarf the $\nu_{3}$ band
should be detectable, as perhaps the first sign of the change taking
place in its atmospheric chemistry.

In May 2000 we used UKIRT and CGS4 to search for this band in a set of L
dwarfs of different spectral types.  As shown in Fig.~4 a broad
absorption feature, which we identify as the Q branch of the $\nu_{3}$
band of CH$_{4}$, was detected in the two latest objects observed,
2MASS~1507 and 2MASS~0825 [11], classified as L5 and L7.5, respectively
[12].  The observed feature is slightly shifted with respect to the
strong Q branch absorption in the earth's atmosphere, due to the higher
temperatures in the dwarfs, which cause the excitation of higher J levels
of the molecule. The extension of the absorption to longer wavelengths
makes the Q branch more readily detectable from the ground. We also
obtained a spectrum of 2MASS~1507 in the K band, finding no evidence for
methane absorption in that band. No K band spectrum of the later object,
2MASS~0825 was obtained.

Models of L dwarf spectra in the 3-4~$\mu$m region (Fig.~5) show the
growth of methane absorption with decreasing effective temperature.  
Methane abundances at each layer of the model atmosphere were calculated
using published chemical equilibrium abundance profiles [13].  The
strength of the methane absorption is weakly dependent on model
assumptions about cloud structure; the models in Fig.~5 are for a clear
atmosphere.  The weakness of the observed methane Q branches in the
spectra of 2MASS~1507 and 2MASS~0825 (Fig.~4)  and the absence of
absorption due to the P and R branches suggest that these dwarfs have
effective temperatures near 1800~K.  This is in general agreement with
temperatures suggested in [14], but higher than the temperature inferred
in [15] for these spectral classes.  Because of methane's
temperature-sensitive chemistry and because methane is expected to be
found well above any condensation clouds, the $\nu_{3}$ band promises to
be an effective indicator of temperature for late L dwarfs.

\section{Classification issues}

The discovery of transition objects and the detection of absorption by
methane in L-type dwarfs bring to the fore several issues pertaining to
classification and terminology.

1. {\it Do the three ``transition objects'' belong to type L or T?} Our
view is that these objects should define the earliest part of the T
sequence, which up to now shows very little spectral variation compared
to that of type L. Moreover, the latter class appears to be running short
of available subtypes in the classification systems proposed at present.

2. {\it However they are classified, what spectral features should be
used in defining the transition sub-types?} As is shown in this paper,
large changes between the spectra of late L-type and classical T-type
objects occur in the 1-2.5~$\mu$m region.  These changes are much greater
than those that occur at shorter wavelengths for the same objects.
Furthermore, because the optical--near infrared colors of ``transition''
dwarfs and classical T dwarfs both are extremely red, they are as or more
tractable to observe in the J, H, and K bands than at optical
wavelengths, Many of them are easily observable in these bands by 4 meter
class telescopes. These characteristics make 1-2.5~$\mu$m the most
attractive interval for defining the classification scheme. Because of
their strong variation with temperature we regard the methane bands at
1.6-1.7~$\mu$m and 2.2-2.4~$\mu$m and the water band at 1.15~$\mu$m (see
Fig.~3) as promising candidates to employ in defining sub-types. Although
the methane bands at 2.2~$\mu$m and longer are stronger that the bands at
1.6-1.7~$\mu$m, contamination of the former by CO and the variable
depression of the continuum peak near 2.1~$\mu$m by the H$_{2}$
pressure-induced dipole absorption (which is more pressure sensitive than
temperature sensitive) may make use of the K band problematic.

3. {\it Should the discovery of CH$_{4}$ at 3.3~$\mu$m in mid and late L
types affect the proposed L and incipient T classification schemes?} We
believe not. In large part the reasons have to do with
practicality. Accurate ground-based measurements in the L band are
considerably more difficult to obtain than similar measurements at shorter
wavelengths. In the L band obtaining signal-to-noise ratios sufficiently
high for purposes of classification will be possible for only a small
fraction of all L-type dwarfs, compared to the much larger fractions that
will be observable spectroscopically at optical and shorter infrared
wavelengths.  Thus we suggest that dwarfs showing the $\nu_{3}$ and
(possibly) longer wavelength fundamental bands of CH$_{4}$, but not the
combination bands in the 1-2.5~$\mu$m region should remain as L-types. 

Although we have suggested that the 3~$\mu$m methane band not enter into
classification schemes, we point out that until a more comprehensive set
of JHK-band spectra are in hand for late L (in currently proposed schemes)
and ``transition'' dwarfs, it probably is premature to define the L-T
boundary. Finally, the detection of methane in L-type objects indicates
that the term ``methane dwarf,'' which previously has been used
interchangeably with ``T dwarf,'' should now be used with caution.

%INDEX%%%%%%%%%%%%%%%%%%%%%%%%%%%%%%%%%%%%%%%%%%%%%%%%%%%%%%%%%%%%%%%
% Please check with the editor of your book whether he plans to
% include a "mutual" subject index - if so, please code your entries
% in the standard syntax. For your own purposes you may print your
% "personal" index by using the following commands:
%
%\clearpage
%\addcontentsline{toc}{section}{Index}
%\flushbottom
%\printindex
%%%%%%%%%%%%%%%%%%%%%%%%%%%%%%%%%%%%%%%%%%%%%%%%%%%%%%%%%%%%%%%%%%%%%

\end{document}